\documentstyle[epsf]{cjp3}
\newcommand{\beas}{\begin{eqnarray*}}
\newcommand{\eeas}{\end{eqnarray*}}
\def\del{\partial}

\begin{document}

\title{Feynman-Schwinger technique in field theories\footnote{Lectures at the 13th Indian-Summer School on Intermediate Energy Physics, {\em Understanding 
the Structure of Hadrons}, Prague, Czech Republic, August 28 - September 1, 
2000.}}

\authori{\c{C}etin \c{S}avkl{\i}}

\addressi{Department of Physics, College of William and Mary, Williamsburg,
          Virginia 23187, USA}

\authorii{}       
\addressii{}
\authoriii{}      
\addressiii{}     

\headtitle{Feynman-Schwinger technique in field theories}
\headauthor{\c{C}etin \c{S}avkl{\i}}

\evidence{A}
\daterec{30 October 2000}    
\cislo{0}  \year{2000}
\setcounter{page}{1}

\maketitle

\begin{abstract}
In these lectures we introduce the Feynman-Schwinger representation 
method for solving nonperturbative problems in field theory. As an 
introduction we first give a brief overview of integral equations and path 
integral methods for solving nonperturbative problems. Then we discuss
the Feynman-Schwinger (FSR) representation method with applications 
to scalar interactions. The FSR approach is a continuum path integral 
integral approach in terms of covariant trajectories of particles.
Using the exact results provided by the FSR approach we test the reliability 
of commonly used approximations for nonperturbative summation of interactions 
for few body systems.

\end{abstract}

\section{Introduction}
Physics research in general is driven by the goal of finding the correct 
Lagrangian for a given system. Once we have a candidate for the correct 
Lagrangian one must be able to relate it to observables. It is clear that 
the question of ``what is the correct Lagrangian?'' is inseparable from 
the question of ``what is the exact prediction dictated by a given 
Lagrangian?'', since we can never be sure that we have the right dynamics if 
we do not know how to calculate the exact result with it.  However, making 
exact predictions using a Lagrangian is not always an easy task. Therefore, 
especially in field theory, one has to make approximations. One common 
approximation is known as perturbation theory. Perturbation theory involves 
making an expansion in the coupling strength of the interaction. It is 
expected to work particularly for small couplings. However, irrespective 
of how small the coupling strength is, it is well known that perturbation 
theory can not explain bound states. This fact can be observed even at the 
level of classical mechanics. Consider the example of a simple harmonic 
oscillator:
\begin{figure}
\begin{center}
\mbox{
   \epsfxsize=1.8in
\epsffile{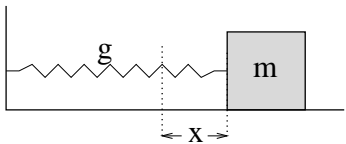}
}
\end{center}
\caption{Simple harmonic oscillator}
\label{oscillator.fig}
\end{figure}
In this simple example the Lagrangian is given by
\bea
L&=&\frac{1}{2} m \dot{x}^2 - \frac{1}{2} g x^2,
\eea
and from the Euler-Lagrange equations the nonperturbative result follows as:
\bea
x(t)&=& A\, {\rm Sin}(wt),\hspace{2cm} w= \sqrt{ \frac{g}{m} }.
\eea
One might express this result as a power series in the coupling strength
\bea
x(t) &=& A \biggl[ wt - \frac{(wt)^3}{3!}+\frac{(wt)^5}{5!}+\cdots\biggr].
\eea
As this expansion shows a perturbative truncation of the above series 
can not produce a bound state. In other words, in order to be able to obtain 
a bound state result one must sum the interactions to all orders. 

The situation in field theory is similar. Bound states in field theory 
are identified by the pole of Green's function.
\begin{figure}
\begin{center}
\mbox{
   \epsfxsize=2.2in
\epsffile{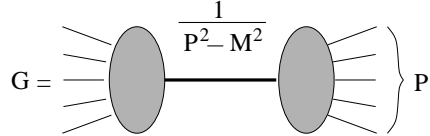}
}
\end{center}
\caption{Bound state mass is determined by the pole of Green's function}
\label{nbody.fig}
\end{figure}
In general the Green's function can be expanded in powers of the coupling 
strength. The question is then: Can we make a truncation in the perturbation 
series at order $g^{2n}$ and still obtain a bound state ? 
\bea
  &&G\stackrel{?}{\simeq} G_0+g^2G_1+g^4G_2+\cdots+g^{2i}\fbox{$G_i$}+\cdots+g^{2n}G_n.
\eea
Since this is a finite series, the bound state singularity could only come 
from individual terms
\bea
  && \fbox{$G_i$}\,\propto\,\frac{1}{P^2-M^2} \,\,\, ?
\eea
However this possibility leads to a contradiction since it implies that 
the dynamics of the bound state is independent of $g$, since $g$ comes only as
an overall factor in front of an individual term $G_i$. The case where more 
than one term is singular leads to the same contradiction. Therefore 
individual $G_i$'s must be nonsingular, and bound state singularity must come 
from an infinite summation of the perturbation series. Situation is similar to
the expansion
\bea
\frac{1}{1-x}=1+x+x^2+x^3+\cdots,
\eea
where the singularity of the left hand side can not be obtained by a finite truncation of the right side. Therefore bound states are always fully 
nonperturbative. 

With the discovery of quantum chromo-dynamics (QCD) nonperturbative 
calculations in field theory have become even more essential. It is known that 
the building blocks of matter, quarks and gluons, only exist in bound states. 
Therefore any reaction that involves quarks will necessarily involve bound 
states in the initial and/or final states. This implies that even at high 
momentum transfers, where QCD is perturbative, formation of quarks into a 
bound state necessitates a nonperturbative treatment. Therefore it is 
essential to develop new methods for doing nonperturbative calculations in 
field theory.

The plan of this lecture is as follows. In the following section a review of
nonperturbative methods in field theory will be given. Later in Sections 3 and
4 the Feynman-Schwinger representation will be introduced through examples.
In particular the emphasis will be on comparing various nonperturbative 
results obtained by different methods. It will be shown with examples that 
nonperturbative calculations are interesting and exact nonperturbative 
results could significantly differ from those obtained by approximate 
nonperturbative methods. In the last section simple perturbation theory results
will be derived using the FSR approach.
 
\section{Nonperturbative methods}

Nonperturbative calculations can be divided into two general categories.
These are {\bf i)} Integral equations,  and {\bf ii)} Path integrals.
In the following two subsections these approaches will be briefly discussed
\subsection{Integral Equations}
Integral equations have been used for a long time to sum interactions 
to all orders with various approximations.~\cite{BETHESALPETER,NAKANISHI,GROSS1,GROSS2,TJON0} In general a complete solution
of field theory to all orders can be provided by an infinite set of 
integral equations relating vertices and propagators of the theory to each 
other. However solving an infinite set of equations is beyond our reach 
and usually integral equations are truncated by various assumptions about
the interaction kernels and vertices. The most commonly used integral 
equations are those that deal with 1, and 2-body problems. Here we give 2 
examples for the 2-body bound state problem. The first example is the 
Bethe-Salpeter equation in the ladder approximation.~\cite{BETHESALPETER}
\begin{figure}
\begin{center}
\mbox{
   \epsfxsize=3.0in
\epsffile{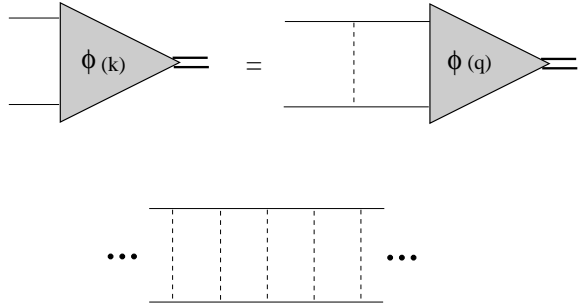} 
}
\end{center}
\caption{Bethe-Salpeter equation in the ladder approximation sums only ladder 
type diagrams}
\label{simplebs4.eps}
\end{figure}
In the ladder approximation the Bethe-Salpeter equation sums ladder diagrams 
to all orders (Fig.~\ref{simplebs4.eps}). Self energies vertex corrections and
 crossed ladder exchanges 
are left out in this approximation.
For the simple case of scalar particles, 
The vertex function $\Phi_{P}(k)$ determining the structure of bound state satisfies an integral equation. For the simple case of scalar particles bound 
state equation can be written as
\bea
\Phi_{P}(k)&=&ig^2\biggl\lmoustache\,\frac{d^4q}{(2\pi)^4}\,G(q-k)\,\frac{\Phi_{P}(q)}{[q^2-M^2][(P-q)^2-m^2]},
\eea
where $G(q-k)$ is the interaction kernel. This equation could be solved using numerical methods to find the bound 
state mass $P$ and the vertex function  $\Phi_{P}(k)$. The vertex function 
is similar to the quantum mechanical wave function. In principle it contains 
all information about the bound state. A serious deficiency of the ladder 
approximation is that it does not have the correct one body limit when 
one of the particles is infinitely heavy.~\cite{GROSS1} This is due to the 
fact that the ladder approximation ignores the crossed ladder type exchanges 
between the particles. As it will be shown by explicit examples in this 
article the crossed ladder exchanges play a crucial role in obtaining the 
correct result not only in the 2-body problem but also in the calculation
of self energies for the 1-body propagators.
 
\begin{figure}
\begin{center}
\mbox{
   \epsfxsize=3.5in
\epsffile{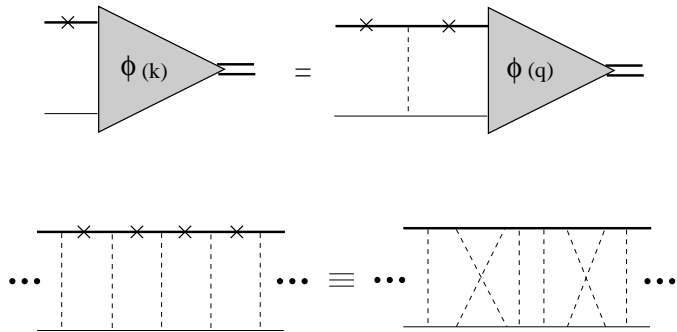} }
\end{center}
\caption{In the limit of infinitely heavy-light systems Gross equation 
effectively sums all ladder and crossed ladder diagrams.}
\label{gross.fig}
\end{figure}

The problem of 1-body limit when one of the particles is infinitely heavy is 
solved by the Gross equation (Fig.~\ref{gross.fig}).~\cite{GROSS1} In the 
Gross equation the 
heavier constituent is constraint to its physical mass shell ($q^2=M^2$)
Putting the heavier constituent on its mass shell and summing only ladder 
diagrams effectively is equivalent to summing all ladder and crossed ladder 
diagrams. When the heavy particle is constraint to its mass shell, the 
bound state equation takes the following form
\bea
\Phi_{P}(\hat{k})&=&g^2\biggl\lmoustache\,\frac{d^3q}{(2\pi)^3\,2E_q}\,G(\hat{q}-\hat{k})\,\frac{\Phi_{P}(\hat{q})}{[(P-\hat{q})^2-m^2]}.
\eea
When bound state of equal (or close) mass particles are under consideration
the Gross equation can be symmetrized by picking up the mass pole contribution
of both particles in different channels. Gross equation is a manifestly 
covariant relativistic equation, and in the nonrelativistic limit Schroedinger
equation is recovered. In the literature this equation has been successfully 
used to analyze relativistic bound states. 

Some general comments on integral equations are as follows: There are a number
of useful features of integral equations. Integral equations are a natural 
extension of nonrelativistic quantum mechanics to field theory. They are a 
practical tool for modeling and doing simple calculations using the tools of field theory. Because of their 
similarity to Schroedinger equation in quantum mechanics, integral equations
in field theory provide an intuitively clear picture of physics. Furthermore
the numerical cost of solving integral equations is negligibly small compared
to methods such as path integrals (to be discussed in the next section).
The last important advantage of integral equations is the fact that they can 
be solved in Minkowski metric. The only other alternative to integral 
equations, path integrals, make use of the Euclidean metric which limits their
applicability. This is an important technical problem -particularly for 
calculations of scattering reactions, form factors and decays- since 
physical particles live in the timelike region.

On the other hand integral equations have some drawbacks. The first problem 
is that they are not exact. Without knowing the exact result it is not 
possible to claim that integral equations are even a good approximation to 
the full theory. The second problem is that integral equations in general 
do not respect the symmetries of the underlying Lagrangian. Except for very 
special approximations, integral equations are not gauge invariant. Therefore 
a more rigorous and systematic approach is needed. This is where path 
integrals play a significant role. In the next section we introduce the 
method of path integrals in quantum mechanics and field theory. 

\subsection{Path Integrals}
Path integrals provide a systematic method for summing interactions to all 
orders. First we begin by giving path integral expressions for quantum 
mechanics. The matrix element for a transition from an initial state $| q_i,t_i\rangle$ to a final state $| q_f,t_f \rangle$ is given by 
\bea
\langle q_f,t_f | q_i,t_i\rangle_H&\equiv&\langle q_f,t_f|\,e^{-iH\,(t_f-t_i)}|q_i,t_i\rangle_S,\\
\langle q_f,t_f|\,e^{-iH\,(t_f-t_i)}|q_i,t_i\rangle &=&\biggl\lmoustache \,{\cal D}q\,e^{\,i\bigl\lmoustache_{t_i}^{t_f} L(q,\dot{q},t)\, dt/\hbar},
\eea
where subscripts H, and S refer to Heisenberg and Schroedinger states. In this
path integral expression all trajectories contribute to the final result with 
equal weight. 
\begin{figure}
\begin{center}
\mbox{
   \epsfxsize=1.7in
\epsffile{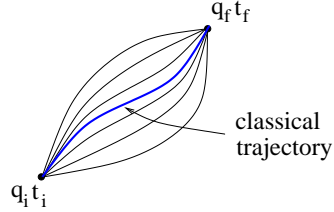}
}
\end{center}
\caption{In quantum mechanics all paths contribute.}
\label{paths.fig}
\end{figure}
Classical limit can be readily obtained by letting $\hbar\rightarrow 0$. In
this limit small variations in the action will lead to large oscillations
and the dominant contribution will come from those trajectories that minimize
the action:
\bea
\delta S &=& \delta \biggl\lmoustache_{q_i}^{q_f} \,L[q,\dot{q},t]\,dt=0.
\eea
The minimization of the action leads to Euler-Lagrange equations of classical 
mechanics:
\bea
\frac{d}{dt}\biggl[\frac{\del L}{\del\dot{q}}\biggr]-\frac{\del L}{\del q}&=&0.
\eea
Keeping the field theory applications in mind we consider the time ordered
products of operators. The time ordered product of operators can also be 
expressed in the form of a path integral:
\beas
\langle q_f,t_f |T(\hat{q}(t_1)\hat{q}(t_2)\cdots \hat{q}(t_n)|q_i,t_i\rangle&=&\biggl\lmoustache{\cal D}q\,q(t_1)\,q(t_2)\cdots q(t_n)\,{\rm exp}\biggl[i\int_{t_i}^{t_f}dt\,L(q,\dot{q})\biggr].
\eeas
In field theory applications the ground state expectation values of time 
ordered products are particularly needed. In order to show how the 
ground state expectation value can be obtain let us introduce a complete set 
of energy eigenstates into the 2-point Green's function
\bea
\langle q_f,t |T[\hat{q}(t_1)\hat{q}(t_2)]|q_i,-t\rangle&=&\sum_{n,m=0}^\infty\langle q_f,t_f |n \rangle\langle n| T[\hat{q}(t_1)\hat{q}(t_2)] |m\rangle\langle m|q_i,t_i\rangle.
\eea
Noting that
\bea
\langle q,t|n\rangle&=&\psi_n(q)e^{-iE_n t},
\eea
one obtains
\beas
\langle q_f,t |T[\hat{q}(t_1)\hat{q}(t_2)]|q_i,-t\rangle&=&\sum_{n,m=0}^\infty {\rm e}^{-i(E_n+E_m)t}\psi_n(q_f)\psi_m(q_i)\langle n| T[\hat{q}(t_1)\hat{q}(t_2)]|m\rangle.
\eeas
As $t\rightarrow -i\infty$ the ground state $(n=0)$ will give the dominant contribution,
\bea
\lim_{t\rightarrow\infty}\langle q_f,t |T[\hat{q}(t_1)\hat{q}(t_2)]|q_i,-t\rangle&\simeq&\underbrace{{\rm e}^{-2E_0t}\psi_0(q_f)\psi_0(q_i)}_{\langle q_f, t|q_i,-t\rangle} \langle 0| T[\hat{q}(t_1)\hat{q}(t_2)]|0\rangle.
\eea
Therefore the ground state expectation value is obtained as
\bea
\langle 0| T[\hat{q}(t_1)\hat{q}(t_2)]|0\rangle&=&\frac{\langle q_f,\infty |T[\hat{q}(t_1)\hat{q}(t_2)]|q_i,-\infty\rangle}{\langle q_f, \infty|q_i,-\infty\rangle}.
\eea
In general the ground state expectation value of the time ordered products
of an arbitrary number of operators can be written as
\beas
\langle 0| T[\hat{q}(t_1)\hat{q}(t_2)\cdots \hat{q}(t_n)]|0\rangle&=&\frac{\biggl\lmoustache{\cal D}q\,q(t_1)\,q(t_2)\cdots q(t_n)\,{\rm exp}\biggl[i\int_{-\infty}^{\infty}dt\,L(q,\dot{q})\biggr]}{\biggl\lmoustache{\cal D}q\,{\rm exp}\biggl[i\int_{-\infty}^{\infty}dt\,L(q,\dot{q})\biggr]}.
\eeas
In going from quantum mechanics to quantum field theory we recognize that a 
field is an infinite array of coordinates, such as an infinite array of 
coupled oscillators, and particles are associated with the normal modes of 
oscillators. Therefore the quantum mechanical derivation can be generalized by
replacing
\bea
q(t)\rightarrow\phi({\bf x}_1,t),\phi({\bf x}_2,t),\cdots ,\phi({\bf x}_n,t),\cdots,
\eea
where $\phi({\bf x}_i,t)$ is the displacement of the oscillator at $x_i$.
\begin{figure}
\begin{center}
\mbox{
   \epsfxsize=2.0in
\epsffile{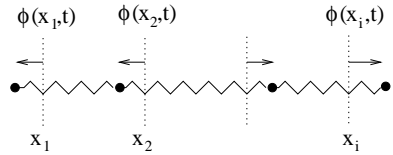}
}
\caption{Field as the coordinates of an infinite array of coupled oscillators}
\end{center}
\label{field.fig}
\end{figure}
The generalized path integral in terms of fields is given by:
\bea
\langle\phi({\bf x},t_f)|\,{\rm exp}\biggl[-iH\,(t_f-t_i)\biggr]|\phi({\bf x},t_i)\rangle&=&\biggl\lmoustache \,{\cal D}\phi\,{\rm exp}
\biggl[\,i\int_{t_i}^{t_f} \,d^4xL\,\biggr],
\eea
where
\bea
{\cal D}\phi=\Pi_i d\Phi(x_i),
\eea
represents a sum over all possible field configurations. The step of going 
from particle trajectories $q(t)$ to fields $\phi({\bf x},t)$ dramatically 
increases the dimensionality of the problem. While quantum mechanical path 
integral sums all possible trajectories (line configuration), field 
theoretical path integral sums all possible field configurations (volume 
configuration).

The Green's function in field theory is given by the path integral expression: 
\beas
\langle 0| T[\phi(x_1)\phi(x_2)\cdots \phi(x_n)]|0\rangle&=&\frac{\biggl\lmoustache[{\cal D}\phi]\,\phi(x_1)\,\phi(t_2)\cdots \phi(x_n)\,{\rm exp}\biggl[i\int d^4x\,L(x)\biggr]}{\biggl\lmoustache[{\cal D}\phi]\,{\rm exp}\biggl[i\int d^4x\,L(x)\biggr]}.
\eeas
This result can be obtained from a generating function
\bea
\langle 0 |T(\phi(x_1)\phi(x_2)\cdots \phi(x_n)|0\rangle&=&(-i)^n\frac{\delta^nZ[J(x)]}{\delta J(x_1)\delta J(x_2)\cdots \delta J(x_n)}\Bigg\vert_{J=0},
\eea
where
\bea
Z[J(x)]&=&\biggl\lmoustache[{\cal D}\phi]\,{\rm exp}\biggl[i\int d^4x\,[L(x)+J(x)\phi(x)]\biggr].
\eea
While path integrals provide a compact expression for the exact 
nonperturbative result for propagators, evaluation of the path integral 
is a nontrivial task. In general, field theoretical path integrals must 
be evaluated by numerical integration methods, such as Monte-Carlo integration.
The best known numerical integration method is lattice gauge 
theory.~\cite{ROTHE} Lattice gauge theory involves a discretization of 
space-time.  
\begin{figure}
\begin{center}
\mbox{
   \epsfxsize=1.5in
\epsffile{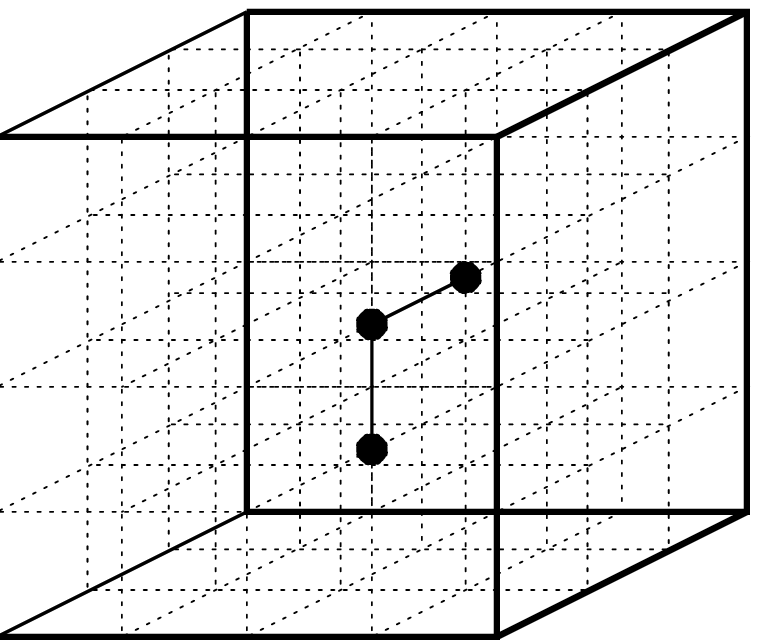}
}
\end{center}
\caption{In lattice gauge theory space-time is discretized within a finite 
box.}
\label{lattice.fig}
\end{figure}
In a discretized lattice particles can be located on the discrete lattice 
sites and exchange fields are represented by links between the sites. After 
discretization the path integral can be performed essentially by a brute force
method. Particularly for QCD lattice gauge theory is currently the only 
method that can produce nonperturbative results starting directly from the 
QCD Lagrangian. However lattice calculations are not without drawbacks.
Discretization of space-time by a cubic lattice violates rotational symmetry. 
In addition the cost of computations critically depend on the size of 
the lattice. Because of this limitation complex applications such as 
calculation of form factors or scattering reactions are beyond the reach of 
current lattice applications. Finally, matter loops are not accounted 
for (quenched approximation) in most lattice calculations due to high 
cost.

In the next section we present a more efficient method of performing path 
integrals in field theory for simple scalar interactions. This method is 
known as Feynman-Schwinger representation(FSR).~\cite{FSR,SIMONOV1,SIMONOV2,SIMONOV3,TACO,BRAMBI,SAVKLI1,SAVKLI2,SAVKLI3} Through applications of the FSR, 
the importantance of {\em exact} nonperturbative calculations will be shown 
with explicit examples.

\section{Feynman-Schwinger representation approach}
The basic idea in the FSR approach is to transform the field theoretical 
path integral such that a quantum mechanical path integral 
in terms of particle trajectories is obtained. In this section we consider 
application of the FSR technique to scalar QED. The Minkowski metric 
expression for the scalar QED Lagrangian in Stueckelberg form is given by
\beas
{\cal L}_{SQED}&=&-m^2\chi^2-\frac{1}{4}F^2+\frac{1}{2}\mu^2A^2-
\lambda\frac{1}{2}(\partial A)^2+(\partial_\mu-ieA_\mu)\chi^*(\partial^\mu+ieA^\mu)\chi,
\eeas
where $A$ represents the gauge field of mass $\mu$, and $\chi$ is 
the charged field of mass $m$. The presence of a mass term for the exchange 
field breaks the gauge invariance. Here the mass term was introduced in order 
to avoid infrared singularities when application to 0+1 dimension is considered
later in the next section. For dimensions larger than n=2 the infrared 
singularity does not exist and therefore the limit $\mu\rightarrow~0$ can be 
safely taken to insure gauge invariance.    

\begin{figure}
\begin{center}
\mbox{
   \epsfxsize=2.0in
\epsffile{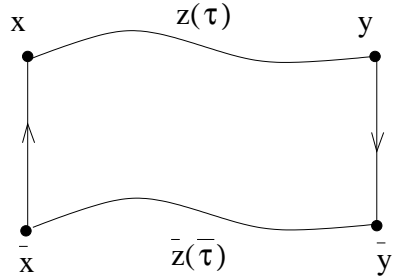} }
\end{center}
\caption{Wilson loop.}
\label{wilsonloop.fig}
\end{figure}

The path integral is to be performed in Euclidean metric. Therefore we perform
a Wick rotation:
\beas
{\rm exp}\biggl[i\int d^4x\,{\cal L}_M\biggr]\longrightarrow {\rm exp}\biggl[-\int d^4x\,{\cal L}_E\biggr].
\eeas
The Wick rotation for coordinates is obtained by
\bea
x_0&\rightarrow& -ix_0\\
\partial_0=\frac{\partial}{\partial x_0}&\rightarrow&i\partial_0.
\eea
The transformation of field $A$ under Wick rotation is found by noting 
that under a gauge transformation:
\bea
A_\mu&\rightarrow&A_\mu+\partial_\mu\Lambda.
\eea
Then, under a Wick rotation:
\bea
A_0&\rightarrow&iA_0.
\eea
The Wick rotated Lagrangian for SQED is given by:
\bea
{\cal L}_{SQED}&=&\chi^*\biggl[m^2-\partial^2-2ieA\partial-ie\partial A+e^2A^2\biggr]\chi+{\cal L}_{A}.
\eea
The exchange field part of the Lagrangian is given by
\bea
{\cal L}_{A}&\equiv&\frac{1}{2}A_\mu(\mu^2g_{\mu\nu}-\lambda\partial_\mu\partial_\nu)A_\nu+\frac{1}{4}F^2,\\
&=&\frac{1}{2}A_\mu[(\mu^2-\Box )g_{\mu\nu}+(1-\lambda)\partial_\mu\partial_\nu]A_\nu.
\eea
We employ the Feynman gauge $\lambda=1$ which yields
\bea
{\cal L}_{A}&=&\frac{1}{2}A_\nu(\mu^2-\Box )A_\nu.
\eea
The two-body Green's function for the transition from an initial state~$\Phi_{i}$ to final state $\Phi_{f}$ is given by
\be
G(y,\bar{y}|x,\bar{x})=N\int {\cal D}\chi^*\int {\cal D}\chi\int {\cal D}A\,
\,\Phi^*_f\Phi_i\,e^{-S_E},
\label{g0.sqed}
\ee
where 
\be
S_E=\int d^4x \,\,{\cal L}_{SQED},
\ee
and a gauge invariant 2-body state $\Phi$ is defined by
\bea
\Phi(x,\bar{x})&=&\chi^*(x)U(x,\bar{x})\chi(\bar{x}).
\eea 
The gauge link $U(x,y)$ which insures gauge invariance of bilinear product of 
fields is defined by  
\begin{equation}
U(x,y)\equiv {\rm exp} \left[-ie\int_x^ydz\,A(z)\right].
\end{equation} 
One can easily see that under a local gauge transformation
\bea
\chi(x) & \rightarrow &{\rm e}^{ie\Lambda(x)}\chi(x)\\
A_\mu(x) & \rightarrow & A_\mu(x)+\partial_\mu\Lambda(x),
\eea
$\Phi_{i}(x,\bar{x})$ remains gauge invariant
\bea
\Phi(x,\bar{x})&\rightarrow&{\rm exp}\biggl[\underbrace{-ie\Lambda(x)+ie\Lambda(\bar{x})-ie\biggl\lmoustache_x^{\bar{x}}dz_\mu\,\partial_\mu\Lambda}_0\biggr]\chi^*(x)U(x,\bar{x})\chi(\bar{x})\\
&=&\Phi(x,\bar{x}).
\eea
Performing path integrals over $\chi$ and $\chi^*$ fields in Eq.~\ref{g0.sqed}
one finds
\bea
G(y,\bar{y}|x,\bar{x})&=&N\bigg\lmoustache {\cal D}A\, ({\rm det}S)\,\,U(x,\bar{x})U^*(y,\bar{y})S(x,y)S(\bar{x},\bar{y})\,e^{-S[A]},
\label{green0}
\eea
where the interacting 1-body propagator $S(x,y)$ is defined by
\bea
S(x,y)&\equiv& \langle y\,|\,\frac{1}{m^2+H(\hat{z},\hat{p})}\,|\,x\rangle\label{s0},\\
H(\hat{z},\hat{p})&\equiv& (\hat{p}+ieA(\hat{z}))^2\label{hami}.
\eea
The Green's function Eq.~\ref{green0} in principle includes contributions 
coming from all possible interactions. The determinant in Eq.~\ref{green0} 
accounts for matter loops and in the quenched approximation it is set equal to
one (${\rm det} S\rightarrow 1$).   
\begin{figure}
\begin{center}
\mbox{
   \epsfxsize=2.4in
\epsffile{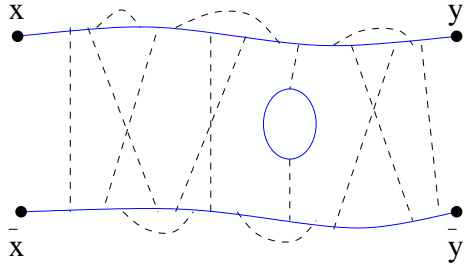}
}
\end{center}
\caption{Various interactions in presence of a matter loop}
\label{unquenched.fig}
\end{figure}
Analytical calculation of the path integral over gauge field $A$ in Eq.~\ref{green0} seems difficult due to nontrivial $A$ dependence in $S(x,y)$. In more complicated theories such as QCD integration of gauge field integral,
as far as we know, is not analytically doable. Therefore, in QCD, the only 
option is to do the gauge field path integral by using a brute force method on
a discretized space-time lattice. However for the simple scalar QED 
interaction under consideration it is in fact possible to go further
and eliminate the path integral over field A. In order to be able to carry out
the remaining path integral over the exchange field $A$ it is desirable to 
represent the interacting propagator in the form of an exponential. This can 
be achieved by using a Feynman representation for the interacting propagator. 
The first step involves the exponentiation of the denominator in Eq.~\ref{s0}:
\bea
S(x,y)&=&\int_0^{\infty}ds\,\, \,e^{-s m^2}\,\langle y\,|\,{\rm exp}[-sH]\,|\,x\rangle,
\label{sqeds}
\eea
This expression is similar to a quantum mechanical propagator where $s=it$, 
and H is a covariant Hamiltonian in terms of 4-vector momentum and coordinates.
It is known that one can use a path integral representation for quantum 
mechanical propagators. A covariant Lagrangian can easily be obtained from
the Hamiltonian 
\bea
&&H(\hat{z},\hat{p})=(\hat{p}+ieA(\hat{z}))^2,\,\,\Longrightarrow\,\,L(z,\dot{z})=\frac{\dot{z}^2}{4}-ie\dot{z}A(z).
\eea
Using this Lagrangian a path integral representation for the interacting 
propagator can be constructed
\bea
S(x,y)&=&\bigg\lmoustache_0^\infty ds\,\bigg\lmoustache\,{\cal D}z\,\,{\rm exp}\biggl[-sm^2-\frac{1}{4}\int_0^s d\tau\, \dot{z}^2(\tau)-ie\int_0^s d\tau\, \dot{z}A(z(\tau))\,\,\biggr],
\eea
where the boundary conditions are given by $z(0)=x$, $z(s)=y$. This 
representation allows one to perform the remaining path integral over the 
exchange field $A$. The final result for the two-body propagator 
involves a quantum mechanical path integral that sums up contributions coming 
from all possible {\em trajectories} of {\em particles}
\bea
G&=&-\bigg\lmoustache_0^\infty 
ds\,\, 
\bigg\lmoustache_0^\infty 
d\bar{s}\,\, \bigg\lmoustache\, 
({\cal D}z)_{xy}\,
\bigg\lmoustache\, 
({\cal D}\bar{z})_{\bar{x}\bar{y}}\,\, 
e^{ -K[z,s]-K[\bar{z},\bar{s}] }    
\langle W(C)\rangle,
\label{gfin}
\eea
where the kinetic term is defined by
\bea
K[z,s]&=&m^2s+\frac{1}{4s}\int_0^1 d\tau \,\dot{z}^2(\tau),
\eea
and the Wilson loop average $\langle W(C)\rangle$ is given by
\bea
\langle W(C)\rangle&\equiv&\int {\cal D}A\,{\rm exp}\biggl[-ie\oint_Cdz\,A(z)-\frac{1}{2}\int d^4z\,A(z)(\mu^2-\del^2)A(z)\biggr],
\eea
where the contour of integration $C$ (Fig.~\ref{wilsonloop.fig}) follows a clockwise trajectory $x\rightarrow y\rightarrow\bar{y}\rightarrow \bar{x}\rightarrow x$ as parameters $\tau$, and 
$\bar{\tau}$ are varied from 0 to $1$. 
The $A$ integration in the definition of the Wilson loop average is 
of standard gaussian form and can be easily performed to obtain
\bea
\langle W(C)\rangle&=&{\rm exp}\,\biggl[-\frac{e^2}{2}\int_C\,dz_\mu\,\int_C\,d\bar{z}_\nu \,\Delta_{\mu\nu}(z-\bar{z},\mu)\biggl],\label{v.sqed}\\
\Delta_{\mu\nu}(x,\mu)&=& g_{\mu\nu}\bigg\lmoustache \frac{d^4p}{(2\pi)^4}\frac{e^{ip x}}{p^2+\mu^2},
\eea
The self energy and the exchange interaction contributions, which are embedded
in expression~\ref{v.sqed}, have different signs. This follows from the fact 
that particles forming the two body bound state carry opposite charges. By 
the result given in Eq.~\ref{gfin} and Eq.~\ref{v.sqed} path integration 
expression involving {\em fields} has been transformed into a path integral 
representation involving {\em trajectories} of particles. 
The bound state spectrum can be determined from the spectral decomposition of the two body Green's function
\be
G(T)=\sum_{n=0}^{\infty}c_ne^{-m_nT},
\ee
where T is defined as the average time between the initial and final states
\be
T\equiv\frac{1}{2}(y_4+\bar{y}_4-x_4-\bar{x}_4).
\label{tdef}
\ee
In the limit of large $T$, the ground state mass is given by
\begin{equation}
m_0=\lim_{T\rightarrow\infty}-\frac{d}{dT} {\rm ln}[G(T)]=\frac{\int {\cal D}Z S'[Z]e^{-S[Z]}}{\int {\cal D}Z e^{-S[Z]}}.
\label{groundstate}
\end{equation}

\subsection{Application to 0+1 dimensions}
Massive scalar QED in 0+1 dimension is a simple interaction that enables one 
to obtain a fully analytical result for the dressed and bound state masses 
within the FSR approach. In this section we compare the self energy result 
obtained by approximate methods with the full result obtained from the 
Feynman-Schwinger representation. In 0 space + 1 time dimension there is no 
continuum spectrum, and only bound states exist.

In general Wilson loop average depends on the trajectory of particles:
\bea
\langle W(C)\rangle&=&{\rm exp}\,\biggl[-\frac{e^2}{2}\int_C\,dz\,\int_C\,d\bar{z} \,\Delta(z-\bar{z},\mu)\biggl].
\eea
In 0+1 dimensions Wilson loop integral is essentially a line integral and
all trajectories contribute equally. Therefore in 0+1 dimension Wilson loop
average does not depend on the shape of trajectory
\bea
{\langle W(C)\rangle}&=&{\rm exp}\,\biggl[-\frac{e^2}{2}\int_0^T\,dz\,\int_0^T\,d\bar{z} \,\Delta(z-\bar{z},\mu)\biggl].
\label{wil1}
\eea
Notice that in 0+1d all trajectories lie on a straight line. Possible 
variations can only come from those trajectories which fold onto themselves.
However the contribution of the folded sections of trajectories identically
vanish. In addition, contribution of matter loops to the Wilson loop average
is identically zero. The typical loop contribution in 1-dimension can be 
written as
\bea
\int_0^Tdz\,\biggl(\int_{z_i}^{z_f}dz'+\int_{z_f}^{z_i} dz'\biggr)\Delta(z-z')&=&0.
\eea
The vanishing of matter loop contribution implies that the quenched 
calculations give exact results. Therefore massive SQED in 0+1 dimension is a 
remarkably simple example where we can compare {\em exact analytic} solutions
of field theory with various approximate nonperturbative methods. Furthermore,
analytical results in 0+1d provide a test case for the numerical routines
which are used in higher dimensions.   

In 0+1d the interaction kernel $\Delta(z-\bar{z},\mu)$ is given by
\bea
\Delta(z-\bar{z},\mu)&=&\frac{e^{-\mu|z-\bar{z}|}}{2\mu}.
\eea
Using this kernel the Wilson loop average Eq.~\ref{wil1} is calculated exactly:
\bea
{\langle W(C)\rangle}&=&{\rm exp}\,\biggl[-\frac{e^2T}{2\mu^2}\biggl(1-\frac{1-e^{-\mu T}}{\mu T}\biggr)\biggr].
\label{wil2}
\eea
Remaining integrals over $ds$ and ${\cal D}z$ in Eq.~\ref{gfin} provide 
the free particle exponential fall of $e^{-mT}$  at large times. Therefore 
the spectrum of the Green's function can be trivially calculated using 
Eqs.~\ref{gfin},~\ref{groundstate}, and~\ref{wil2}. 
Exact analytic FSR results for bound state masses are as follows:
\bea
&&{\rm 1-body :} \,\,\,\,\,\,M_1=m+\frac{e^2}{2\mu},\\
&&{\rm 2-body :} \,\,\,\,\,\,M_2=2m,\\
&&{\rm n-body :} \,\,\,\,\,\,M_n=n\,m-n\,(n-2)\,\frac{e^2}{2\mu}.
\eea
Bound state mass results given above include the self energy contributions.
The n-body result shows that for any given coupling strength $e$, there is an 
upper limit in particle number $n$ beyond which results become unstable. 
Dynamics of the two body bound state mass generation is similar to pion mass 
generation. Self energies of particles exactly cancel the binding energy 
to produce a bound state mass that is proportional to the current masses
\bea
(m+\frac{e^2}{2\mu})+(m+\frac{e^2}{2\mu})-\frac{e^2}{\mu}=2m.
\eea
Exact massive SQED results share some common features with QCD in 1+1 d~\cite{THOFT},
\begin{itemize}
  \item $\mu\rightarrow 0\Longrightarrow M_1\rightarrow \infty$, where $\mu$
plays the role of an infrared cut-off.
  \item 2-body bound state mass $M_2$ is independent of $\mu$. 
  \item When ``cut-off'' $\mu\rightarrow 0 \Longrightarrow$  no bound states 
for $n>2$.
  \item When current masses vanishes  $m\rightarrow 0 \Longrightarrow M_2=0$
``chiral limit''
\end{itemize}
This toy model provides a possible test case also for the lattice gauge 
theory calculations.

\subsection{Comparison of the exact FSR results with approximate 
nonperturbative methods}

In this section we take a closer look at 1-body mass pole calculations.
Popular methods frequently used in finding the dressed mass of a 
particle is to do a simple bubble summation or solve the 1-body Dyson-Schwinger
equation in rainbow approximation. It is interesting to compare results 
given by the bubble summation and the Dyson-Schwinger with the exact FSR 
result.  
Below we first give a quick overview of how dressed masses can be obtained in 
bubble summation and the Dyson-Schwinger equation approaches.

\begin{figure}
\begin{center}
\mbox{
   \epsfxsize=2.8in
\epsffile{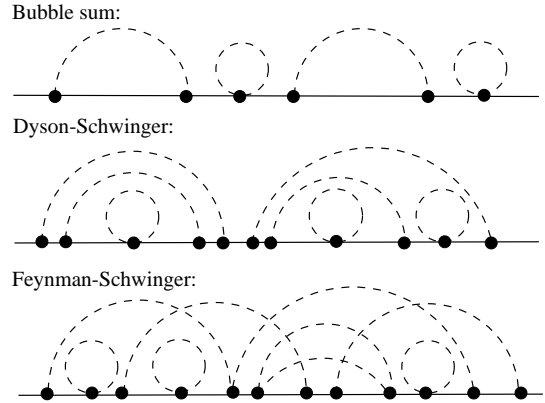} }
\end{center}
\caption{Various interactions included in each approach are shown.}
\label{allthree.fig}
\end{figure}
The simple bubble summation involves a summation of all bubble diagrams (Fig.~\ref{allthree.fig}) to all 
orders. The dressed propagator is given by 
\bea
\Delta_d(p)&=&\frac{1}{p^2+m^2+\Sigma(p)}.
\eea
The dressed mass M is determined from the self energy using
\bea
M=\sqrt{m^2+\Sigma(iM)}.
\label{mdre}
\eea
The self energy for the simple bubble sum is given by
\bea
 \Sigma(p)&=&-e^2\bigg\lmoustache_{-\infty}^{\infty}
  \frac{dk}{2\pi}\frac{1}{(k^2+\mu^2)}\bigg\{\frac{(2p-k)^2}{[(p-k)^2+m^2]}-1\bigg\}.
\eea
The self energy integral in this case is trivial and can be performed 
analytically, and the dressed mass is determined from Eq.~\ref{mdre}

The rainbow Dyson-Schwinger equation sums more diagrams than the simple 
bubble summation (Fig.~\ref{allthree.fig}). The self energy of the 
rainbow Dyson-Schwinger equation involves a momentum dependent mass.
\bea
\Sigma(p)&=&-e^2\bigg\lmoustache_{-\infty}^{\infty}
\frac{dk}{2\pi}\frac{1}{(k^2+\mu^2)}\bigg\{\frac{(2p-k)^2}{[(p-k)^2+\underbrace{m^2+\Sigma(k)}]}-1\bigg\}.
\label{selfe}
\eea
\begin{figure}[b]
\begin{center}
\mbox{
   \epsfxsize=3.9in
\epsffile{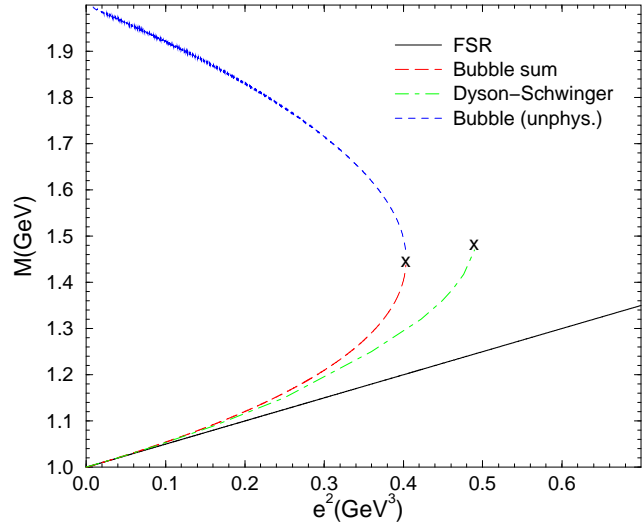} }
\end{center}
\caption{The function $M(e^2)$ calculated by the FSR approach, the
Dyson-Schwinger equation in the rainbow approximation, and the bubble summation for values of 
$m=\mu=1$ GeV. While the exact result is always real, the rainbow DSE and the 
bubble summation results become complex beyond a critical coupling.}
\label{mvsg2.sqed.collide.fig}
\end{figure}
In this case the self energy is nontrivial and it must be determined by
a numerical solution of Eq.~\ref{selfe}. The dressed mass is determined 
by the logarithmic derivative of the dressed propagator in coordinate space  
\bea
M&=&-\lim_{T\rightarrow \infty}\frac{d}{dT}\,\, {\rm log}[\,\Delta_d(t)\,].
\eea
The type of diagrams summed by each method is shown Fig.~\ref{allthree.fig}.
Note that the matter loops do not give any contribution as explained earlier.
Results obtained by these three methods are shown in 
Fig.~\ref{mvsg2.sqed.collide.fig}.
It is interesting to note that the simple bubble summation and the rainbow
Dyson-Schwinger results display similar behavior. While the exact result 
provided by the FSR linearly increases for all coupling strengths, both the 
simple bubble summation  and the rainbow Dyson-Schwinger results come to a 
critical point beyond which solutions for the dressed masses become complex. {\em This 
example very clearly shows that conclusions about the mass poles of propagators
based on approximate methods such as rainbow Dyson-Schwinger equation can 
be misleading}. 
\begin{figure}
\begin{center}
\mbox{
   \epsfxsize=3.2in
\epsffile{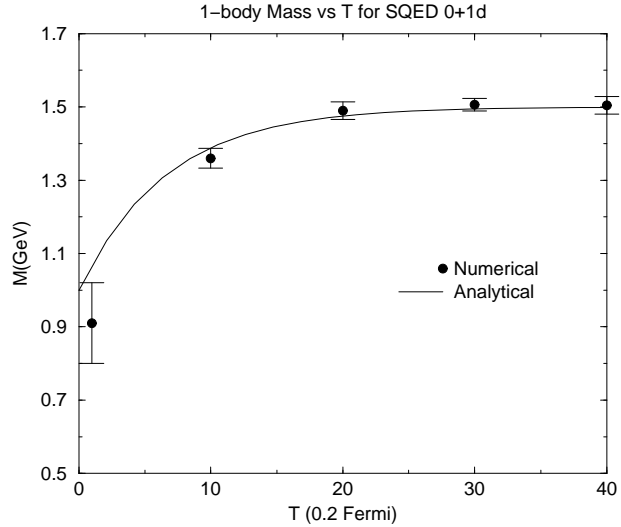} }
\end{center}
\caption{As a test of the numerical methods the numerical Monte-Carlo result 
for 1-body dressed mass in SQED is compared with the analytical result.}
\label{mvst.1b.1d.fig}
\end{figure}

The simplicity of massive SQED in 0+1d also provides an excellent opportunity 
to test the numerical Monte-Carlo integration methods which are normally 
needed at higher dimensions. Therefore as a test of the numerical methods 
used for the remainder of applications in this work, we demonstrate in 
Fig.~\ref{mvst.1b.1d.fig} that the numerical Monte-Carlo results for 1-body
dressed mass in SQED correctly reproduce the analytical result. In 
Fig.~\ref{mvst.1b.1d.fig} the time dependence of the dressed mass given by 
Eq.~\ref{groundstate} is shown. Parameters used for this plot are $\mu=0.15$ 
Gev, and $e=0.15$ $GeV^{1.5}$. 

In the next section we consider the application of the FSR approach to 
scalar $\chi^2\phi$ interaction
\section{Scalar $\chi^2\phi$ interaction with the FSR approach}
\begin{figure}
\begin{center}
\mbox{
   \epsfxsize=3.0in
\epsffile{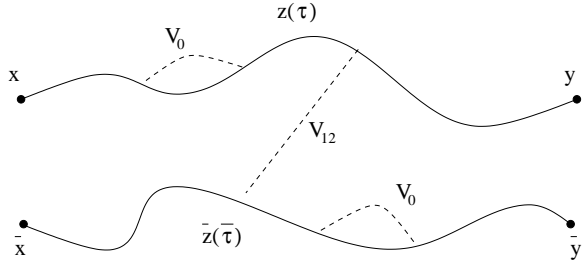} }
\end{center}
\caption{Sample trajectories with self and exchange interactions.}
\label{trajectory.fig}
\end{figure}

\begin{figure}
\begin{center}
\mbox{
   \epsfxsize=2.2in
\epsffile{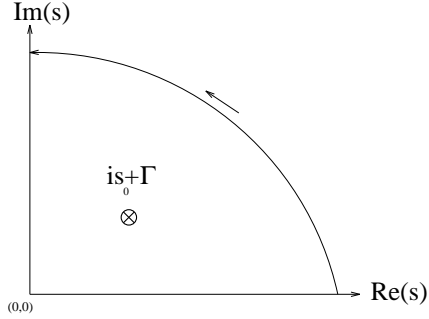} }
\end{center}
\caption{Wick rotation in $s$ integration}
\label{wick.fig}
\end{figure}

\begin{figure}
\begin{center}
\mbox{
   \epsfxsize=3.2in
\epsffile{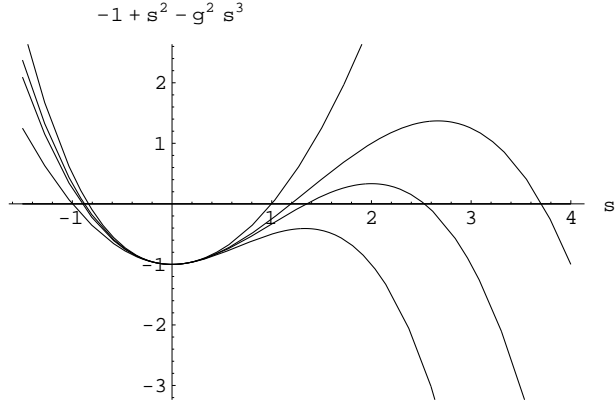} }
\end{center}
\caption{As coupling strength is increased the stationary point disappears.}
\label{gcritical.fig}
\end{figure}

We consider the theory of charged scalar particles $\chi$ of mass $m$ 
interacting through the exchange of a neutral scalar particle $\phi$ of mass 
$\mu$. The Euclidean Lagrangian for this theory is given by 
\bea
{\cal L}&=&\chi^*\bigl[m^2-\del^2+g\phi\bigr]\chi+\frac{1}{2}\,\phi(\mu^2-\del^2)\phi.
\eea
The 2-body propagator for the transition from the initial state
$\Phi_{i}=\chi^*(x)\chi(\bar{x})$ to final state $\Phi_{f}=\chi^*(y)\chi(\bar{y})$ is given by
\bea
&&G(y,\bar{y}|x,\bar{x})=N\bigg\lmoustache {\cal D}\chi^*\bigg\lmoustache {\cal D}\chi\bigg\lmoustache {\cal D}A\,\,\,\Phi^*_f\,\Phi_i\,\,{\rm exp}\biggl[-\int d^4x\,{\cal L}\biggr].
\eea
After the usual integration of matter fields is done the Green's function 
reduces to
\bea
G(y,\bar{y}|x,\bar{x})&=&N\bigg\lmoustache {\cal D}\phi\, ({\rm det}S)\,\,S(x,y)S(\bar{x},\bar{y})\,e^{-S[\phi]}.
\eea
As in the case of scalar QED we employ the quenched approximation: ${\rm det} S\rightarrow 1$. The interacting propagator $S(x,y)$ is defined as
\bea
S(x,y)&\equiv &\langle y\,|\,\frac{1}{m^2+H(\hat{z},\hat{p})}\,|\,x\rangle\\
H(\hat{z},\hat{p})&\equiv& \hat{p}^2-g\phi(\hat{z}).
\label{hphi3}
\eea
We exponentiate the denominator by introducing an $s$ integration along the {\em imaginary} axis with an $\epsilon$ prescription 
\bea
S(x,y)&=&\int_0^{i\infty}ds\,\, \,e^{-s (m^2+i\epsilon)}\,\langle y\,|
\,{\rm exp}[-sH]\,|\,x\rangle.
\eea
This representation should be compared with the representation used earlier 
in SQED Eq.~\ref{sqeds}. Here the integration is done along the imaginary axis
because $H$ is not positive definite. Again a quantum mechanical path integral 
representation can be constructed by recognizing that Lagrangian corresponding
to $H$ Eq.~\ref{hphi3} is given by 
\bea
L(z,\dot{z})=\frac{\dot{z}^2}{4}+g\phi(z).
\eea
The path integral representation for the interacting propagator is
\beas
S(x,y)&=&-i\bigg\lmoustache_0^\infty ds\,\bigg\lmoustache\,{\cal D}z\,\,{\rm exp}\biggl[is(m^2+i\epsilon)-\frac{i}{4}\int_0^s d\tau\, \dot{z}^2(\tau)+ig\int_0^s d\tau\,\phi(z(\tau))\,\,\biggr].
\eeas
This representation allows the elimination of the integral over exchange 
field $\phi$. The 2-body propagator reduces to 
\bea
G&=&-\bigg\lmoustache_0^\infty ds\,\, \bigg\lmoustache_0^\infty d\bar{s}\,\, \bigg\lmoustache\, ({\cal D}z)_{xy}\,\bigg\lmoustache\, 
({\cal D}\bar{z})_{\bar{x}\bar{y}}\,\,e^{iK[z,s]+iK[\bar{z},\bar{s}]} I_\phi,
\eea
where mass and kinetic term is given by
\bea
K[z,s]&=&(m^2+i\epsilon)s-\frac{1}{4s}\int_0^1 d\tau \,\dot{z}^2(\tau).
\eea
The field integration $I_\phi$ is a standard gaussian integration
\bea
I_\phi &\equiv&\biggl\lmoustache 
{\cal D}\phi\,{\rm exp}\biggl[+ig\biggl( \int_0^s d\tau\,\phi(z(\tau)) +\int_0^{\bar{s}} d\bar{\tau}\,\phi(\bar{z}(\bar{\tau})) \biggr)  - S[\phi] \biggr]\\
&\equiv& {\rm exp}\biggl(-V_{0}[z,s]-2\,V_{12}[z,\bar{z},s,\bar{s}]-V_{0}[\bar{z},\bar{s}]\biggr),
\eea
where $V_{0}$ and $V_{12}$ (self and exchange energy contributions in 
Fig.~\ref{trajectory.fig}) are defined by
\bea
V_{0}[z,s]&=&\frac{g^2}{2}\,\,s^2\,\,\bigg\lmoustache_0^1d\tau\bigg\lmoustache_0^1d\tau' \,\Delta(z(\tau)-z(\tau'),\mu),\\
V_{12}[z,\bar{z},s,\bar{s}]&=&\frac{g^2}{2}\,\,s\bar{s}\,\,\bigg\lmoustache_0^1d\tau\bigg\lmoustache_0^1d\bar{\tau}\, \Delta(z(\tau)-\bar{z}(\bar{\tau}),\mu).
\eea
It should be noted that the interaction terms explicitly depend on the $s$ 
variable, which was not the case for SQED. The interaction kernel $\Delta$ is 
given by
\bea
\Delta(x,\mu)&=& \bigg\lmoustache \frac{d^4p}{(2\pi)^4}\frac{e^{ip\cdot x}}{p^2+\mu^2}=\frac{\mu}{4\pi^2|x|}K_1(\mu|x|).
\eea
In order to be able to compute the path integral over trajectories a 
discretization of the path integral is needed
\bea
({\cal D}z)_{xy}&\rightarrow&(N/4\pi s)^{2N}\Pi^{N-1}_{i=1}\bigg\lmoustache d^4z_i.
\label{discreet}
\eea
The s dependence is {\em crucial} for correct normalization. After 
discretization the 1-body propagator takes the following form
\bea
G&=&i\left(\frac{N}{4\pi}\right)^{2N}\bigg\lmoustache \Pi_{i=1}^{N-1}dz_i \bigg\lmoustache_0^\infty \frac{ds}{s^{2N}}\,{\rm exp}\biggl[im^2s-i\frac{k^2}{4s}-s^2v\biggr].
\eea
This is an {\em oscillatory} and {\em regular} integral and it is not 
convenient for Monte-Carlo integration. The origin of the oscillation is the 
fact that $s$ integral was introduced along the imaginary axis,
\bea
&&{\rm Rep.\,\, 1}\,:\,\,\,\,S(x,y)=<y\,|\bigg\lmoustache_0^{-i\infty}ds\, {\rm exp}\biggl[-s(m^2-\del^2+g\phi+i\epsilon)\biggr]|\,x>.
\label{phi3.int.prop.2}
\eea
In earlier works~\cite{SIMONOV2,TACO} a nonoscillatory Feynman-Schwinger 
representation was used,
\bea
&&\hspace{-1.7cm}{\rm Rep.\,\, 2}\,:\,\,\,\,S(x,y)=<y\,|\bigg\lmoustache_0^\infty ds\, {\rm exp}\biggl[-s(m^2-\del^2+g\phi)\biggr]|\,x>.
\eea
Rep. 2 leads to a {\em nonoscillatory} and {\em divergent} result
\bea
G&\propto&\bigg\lmoustache_0^\infty \frac{ds}{s^{2N}}\,{\rm exp}\biggl[-m^2s-\frac{k^2}{4s}+s^2v\biggr],
\eea
and the large $s$ divergence was regulated by a cut-off $\Lambda$. This is not
a satisfactory prescription since it relies on an arbitrary cut-off. Later 
it was shown~\cite{SAVKLI1,SAVKLI3} that the correct procedure is to start with
Rep.1 and make a Wick rotation such that the final result is {\em nonoscillatory}
and {\em regular}.
The implementation of Wick rotation however is nontrivial. Consider
the s-dependent part of the integral for the 1-body propagator
\bea
G&\propto&\bigg\lmoustache_0^\infty \frac{ds}{s^{2N}}\,{\rm exp}\biggl[im^2s-i\frac{k^2}{4s}-s^2\,v\biggr].
\label{gsimple}
\eea
It is clear that a replacement of $s\rightarrow is$ leads to a divergent 
result. The problem with Wick rotation (Fig~\ref{wick.fig}) comes from the 
fact that the $s$ integral is infinite both along the imaginary axis {\em and}
along the contour at infinity. These two infinities cancel to yield a finite 
integral along the real axis.
As $g\rightarrow 0$ the dominant contribution to the $s$ integral in 
Eq.~\ref{gsimple} comes from the stationary point 
\bea
s=is_0\simeq i\frac{k}{2m}.
\eea
Therefore one might suppress the integrand away from the stationary 
point by introducing a damping factor $R$
\bea
R(s,s_0)&\equiv& 1-(s-is_0)^2/\Gamma^2.
\eea
With this factor the integral in Eq.~\ref{gsimple} is modified as
\bea
G&\propto&\bigg\lmoustache_0^\infty \frac{ds}{s^{2N}}\,{\rm exp}\biggl[im^2s-i\frac{k^2}{4s}-\frac{s^2}{R^2(s,s_0)\,v}\biggr].
\label{gcomplex}
\eea
This modification allows us to make a Wick rotation since the contribution 
of the contour at infinity now vanishes. However this procedure relies on the 
fact that there exist a stationary point. It can be seen from the original 
expression Eq.~\ref{gsimple} that this is not always true. According to the 
original integral the stationary point is given by the following equation
\bea
im^2+i\frac{k^2}{4s^2}-2sv&=&0.
\label{ocrit}
\eea
The stationary point is determined by the first intersection of a cubic plot 
with the positive s axis as shown in Fig.~\ref{gcritical.fig}.
The plot in Fig.~\ref{gcritical.fig} shows that as coupling strength is 
increased the curve no longer crosses the positive $s$ axis. Therefore beyond 
a critical coupling strength the stationary point vanishes and mass results 
should be unstable.
Having noted that the original expression Eq.~\ref{gsimple} has a critical 
point, we now turn to perform a Wick rotation on the modified expression 
Eq.~\ref{gcomplex}. Wick rotation in Eq.~\ref{gcomplex} amounts to 
a simple replacement $s\rightarrow is$, and a {\bf nonoscillatory} and 
{\bf regular} integral is found:
\bea
G&\propto&\bigg\lmoustache_0^\infty \frac{ds}{s^{2N}}\,{\rm exp}\biggl[-m^2s-\frac{k^2}{4s}+\frac{s^2}{R^2(is,s_0)}\,v\biggr].
\eea
At first look it seems that the new integral always has a stationary point
determined by the following equation
\bea
-m^2+\frac{k^2}{4s^2}+2sv\frac{1}{R^2(is,s_0)}-s^2v
\frac{(R^2(is,s_0))'}{R^2(is,s_0)}&=&0.
\label{ncrit}
\eea
The key point to remember is that the stationary point we find {\em after}
the Wick rotation should be the same stationary point we had before 
the Wick rotation. This is required to make sure that the physics remains 
the same after the Wick rotation. Therefore self consistency requires that 
the stationary point after the Wick rotation is at $s=is_0$. In that case 
$R(is_0,s_0)=1$, and $(R^2(is_0,s_0))'=0$ and the equation 
determining the critical point Eq.~\ref{ncrit} reduces to the earlier original
form given by Eq.~\ref{ocrit}. Therefore self consistency requirement 
guarantees that the critical point still exists after the Wick rotation. 

The regularization of the ultraviolet singularities are done using 
Pauli-Villars regularization prescription. Pauli-Villars regularization 
is particularly convenient for numerical integration since it only involves 
a change in the interaction kernel
\bea
\Delta(x,\mu)&\longrightarrow& \Delta(x,\mu)-\Delta(x,\alpha\mu).
\eea
\subsection{Numerical applications of $\chi^2\phi$ interaction}
Applications of $\chi^2\phi$ interaction in 3+1d require numerical 
Monte-Carlo integration. First step is the discretization of particle
trajectories,
\begin{figure}
\begin{center}
\mbox{
   \epsfxsize=2.2in
\epsffile{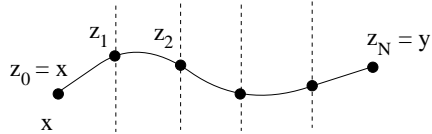} }
\end{center}
\caption{Number of steps a particle takes between initial and 
final coordinates is discretized. The space-time is {\em continuous} and there 
are {\em no space-time boundaries}}
\label{discrete.fig}
\end{figure}
where boundary conditions are given by
\bea
z_0=x=(x_1,x_2,x_3,0)\hspace{2cm}z_N=y=(y_1,y_2,y_3,T).
\eea
Discretization employed in the FSR is for {\em the number of steps} a particle
takes between the initial and final states in a 4-d coordinate space. This is 
very different from the discretization employed in lattice gauge theory. 
Contrary to lattice gauge theory, in the FSR approach space-time is 
{\em continuous} and the rotational symmetry is respected. An additional 
benefit is {\em the lack of space-time} boundary. This is an important 
advantage of the FSR approach. The lack of space-time lattice boundary allows 
analysis of arbitrarily large systems using the FSR approach. This feature
provides an opportunity for doing complex applications such as calculation of 
form factors using the FSR approach. 

\begin{figure}[h]
\begin{center}
\mbox{
   \epsfxsize=3.6in
\epsffile{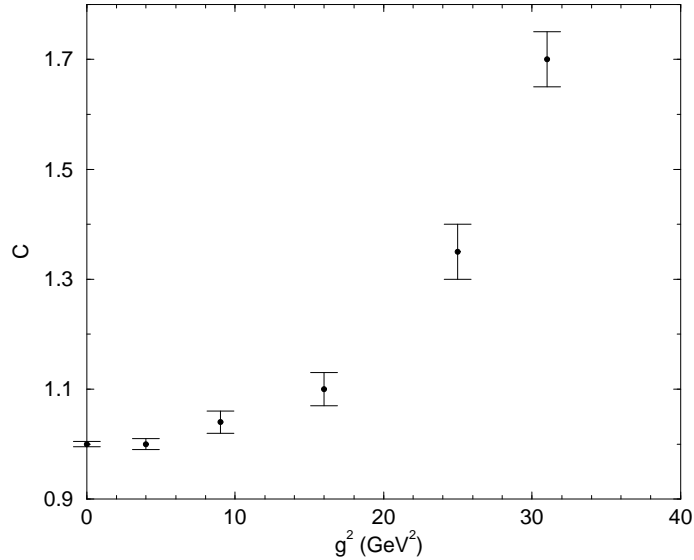} }
\end{center}
\caption{The dependence of the peak of the s-distribution on the coupling 
strength is shown. The peak location is given by $s_0=C T/2m$. Beyond the 
critical coupling strength of $g^2=31 {\rm GeV}^2$ a self consistent 
determination of $C$ is not possible. Therefore beyond the critical coupling strength 1-body mass becomes unstable.}
\label{cvsg2.fig}
\end{figure}
In doing Monte-Carlo sampling we sample {\em trajectories} (lines) rather than
 {\em gauge field configurations} (in a volume). This leads to a significant 
reduction in the numerical cost. 
The ground state mass of the Green's function is obtained using
\begin{equation}
m_0=\frac{\int {\cal D}Z\, S'[Z]e^{-S[Z]}}{\int {\cal D}Z\, e^{-S[Z]}}.
\label{groundstate2}
\end{equation}
Sampling of trajectories is done using the standard Metropolis algorithm. 
Metropolis algorithm insures that configurations sampled are distributed 
according to the weight $e^{-S[Z]}$. In sampling trajectories the final state
(spacial) coordinates of particles are integrated out. Integration over final 
state coordinates puts the system at rest and  projects out the s-wave ground 
state. As trajectories of particles are sampled wave function of the system 
can be determined simply by storing the final state configurations of 
particles in a histogram. 

\begin{figure}
\begin{center}
\mbox{
   \epsfxsize=4.0in
\epsffile{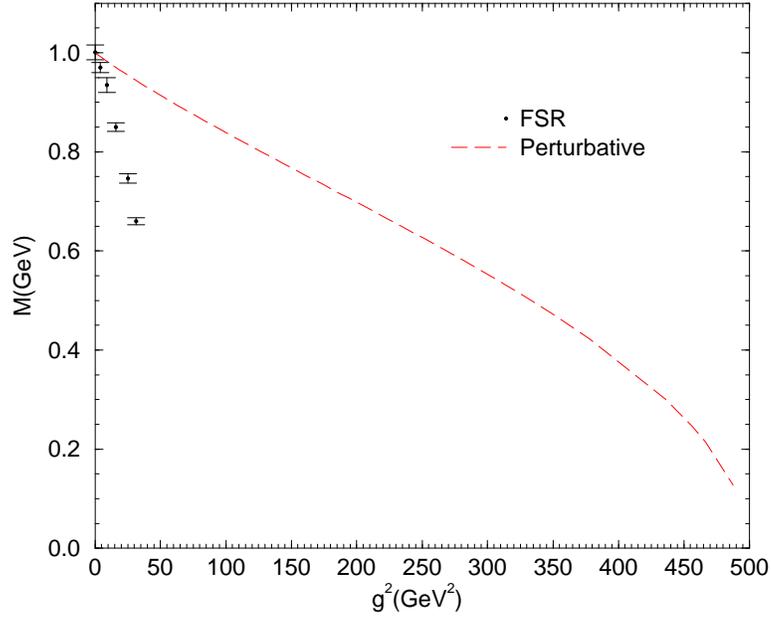} }
\end{center}
\caption{The coupling constant dependence of the 1-body dressed mass is 
shown. Beyond the critical coupling strength of $g^2=31$ ${\rm GeV}^2$ the 
1-body mass becomes unstable. The perturbative bubble summation also has 
a critical point near $g^2=490$ ${\rm GeV}^2$. }
\label{mvsg2.phi3.1b.eps}
\end{figure}
In sampling trajectories the first step is termalization. In order to insure 
that the initial configuration of trajectories has no effect on results 
initial updates are not taken into account. Depending on the dimensionality 
of the problem and the coupling strength the number of initial updates 
neeeded for termalization is of the order of 1000 updates.~\cite{SAVKLI3}  
In order to satisfy self consistency regarding the location of the stationary 
point discussed earlier, the location of the stationary point must be 
determined carefully. The stationary point can be parametrized as 
$s_0=CT/(2m)$, where $T/2m$ is the location of the stationary point 
when the coupling strength $g$ goes to zero. As the coupling strength
is increased the stationary point moves out (see Figs.~\ref{gcritical.fig}, and ~\ref{cvsg2.fig} ) 
and eventually the critical point is reached beyond which there is no 
stationary point. In order to be able to do Monte-Carlo integrations an initial
guess must be made for the location of the stationary point. Self consistency 
is realized by insuring that the peak location of the s distribution in the 
Monte-Carlo integration agrees with the initial guess for the stationary 
point.~\cite{SAVKLI3} In Fig.~\ref{cvsg2.fig} the dependence of the location 
of 
the stationary point on the coupling strength is shown. Fig.~\ref{cvsg2.fig}
shows that beyond the critical point $g^2\simeq 31$ GeV$^2$ $C$ goes to 
infinity implying that there is no stationary point. A similar critical 
behavior was also observed in Refs.~\cite{ROSY} within the context of a variational approach. In Fig.~\ref{mvsg2.phi3.1b.eps} exact 1-body dressed mass results 
are shown for $m_\chi=1$ GeV, $\mu_\phi=0.15$ GeV. Results indicate that the perturbative bubble summation deviates from the quenched FSR result very significantly. These results are all for a Pauli-Villars mass of 3$\mu$.
\begin{figure}
\begin{center}
\mbox{
   \epsfxsize=4.0in
\epsffile{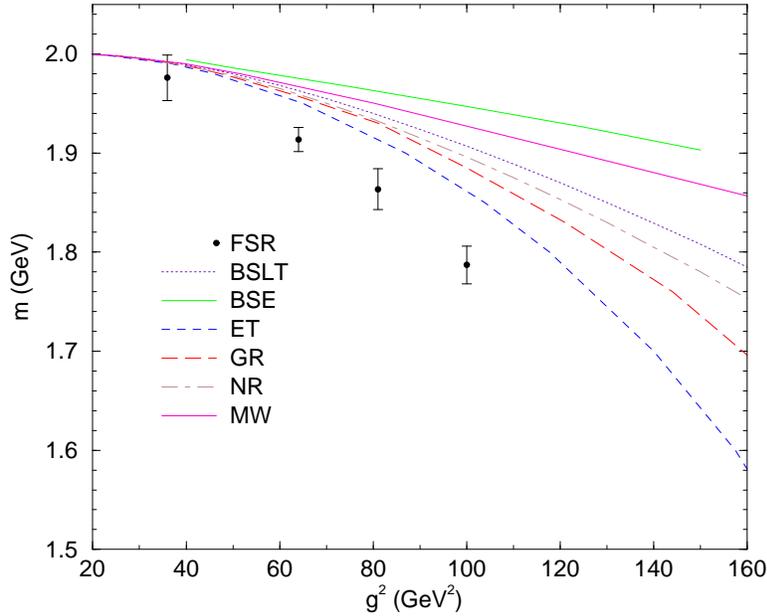} }
\end{center}
\caption{The coupling constant dependence of the 2-body bound state mass is 
shown. Beyond the critical coupling strength of $g^2=100$ GeV$^2$ the 2-body
mass becomes unstable. The Bethe-Salpeter equation in the ladder approximation
gives the lowest binding.}
\label{mvsg2.phi3.2b.fig}
\end{figure}
In Figure~\ref{mvsg2.phi3.2b.fig} we present the 
comparison of the 2-body bound state masses obtained by the FSR with various 
bound state equations. The FSR calculation involves summation of all ladder and crossed ladder diagrams, and excludes the self energy contributions.
According to Figure~\ref{mvsg2.phi3.2b.fig} all bound state equations underbind. Among the manifestly covariant equations the 
Gross equation gives the closest result to the exact calculation obtained by 
the FSR method. This is due to the fact that in the limit of infinitely 
heavy-light systems the Gross equation effectively sums all ladder and 
crossed ladder diagrams. Equal-time equation also produces a strong binding 
but the inclusion of retardation effects pushes the Equal-time results away 
from the exact results (Mandelzweig-Wallace equation~\cite{WALLACE}). In 
particular the 
Bethe-Salpeter equation in the ladder approximation (BSE in 
Figure~\ref{mvsg2.phi3.2b.fig}) gives the lowest binding. Similarly the 
Blankenbecler-Sugar-Logunov-Tavkhelidze equation~\cite{LT,BS} (BSLT) gives a 
very low binding. A comparison of the 
ladder Bethe-Salpeter, Gross, and the FSR results shows that {\em the exchange
of crossed ladder diagrams plays a crucial role.}

\begin{figure}
\begin{center}
\mbox{
   \epsfxsize=3.6in
\epsffile{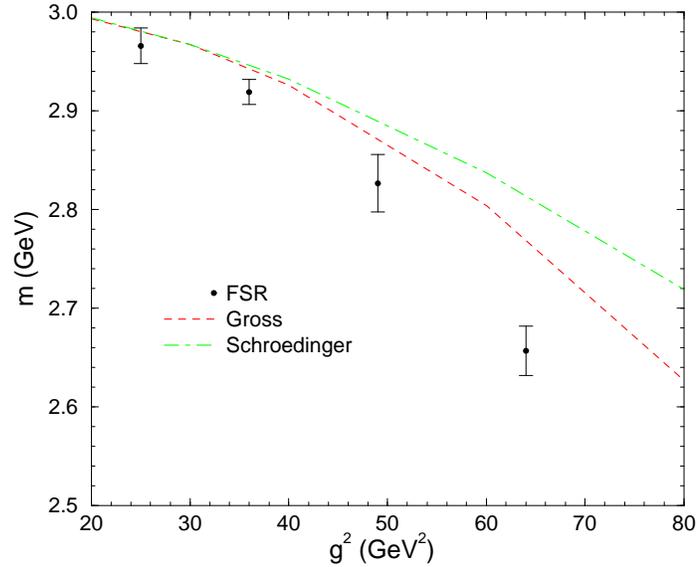} }
\end{center}
\caption{3-body bound state results for 3 equal 
mass particles of mass 1 GeV.}
\label{mvsg2.phi3.3b.fig}
\end{figure}
\begin{figure}
\begin{center}
\mbox{
   \epsfxsize=4.0in
\epsffile{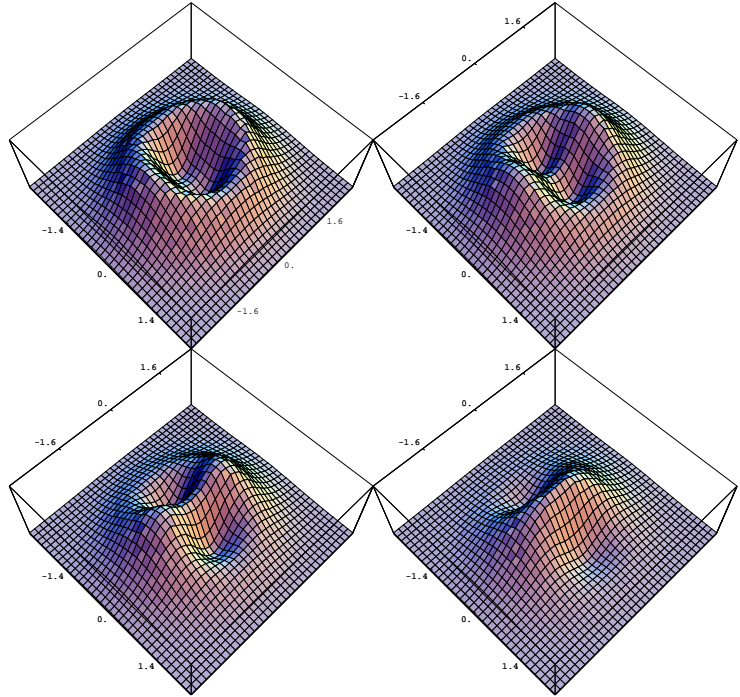} }
\end{center}
\caption{Evolution of the probability distribution for the 3rd particle is
shown as the distance between the two fixed particles is increased. When the 
fixed particles are very close to each other the third particle sees them as a
point particle (the upper left plot). As the fixed particles are separated 
from each other the third particle starts penetrating between them (2nd and 3rd plots), and as the two fixed particles are maximally separated the third particle spends most of its time in between the two fixed particles (the lower 
right plot).}
\label{rdist.surf.3b}
\end{figure}

In Figure~\ref{mvsg2.phi3.3b.fig} the 3-body bound state results for 3 equal 
mass particles of mass 1 GeV is shown. For 3-body case the only available 
results are the Schroedinger and Gross equation results. According to results
presented in Figure~\ref{mvsg2.phi3.3b.fig} bound state equations underbind 
for the 3-body case too. Gross equation gives the closest result to the exact 
FSR result. Determination of the wavefunction of bound states is done by 
keeping the final state configurations of particles in a histogram. For 
example, for a 3-body 
bound state system, the probability distribution of the third particle for a 
given configuration of first and second particles is shown in 
Fig.~\ref{rdist.surf.3b}. In the first plot of 
Fig.~\ref{rdist.surf.3b} two fixed particles are very close to each other such
that the third particle sees them as a point particle. However as the fixed 
particles are separated from each other the third particle starts having 
a nonzero probability of being in between the two fixed particles (second
and third plots of Fig.~\ref{rdist.surf.3b}). Eventually when the 
two fixed particles are kept away from each other the third particle has 
a nonzero probability distribution only at the origin (the last plot 
shown in Fig.~\ref{rdist.surf.3b}).
Until this point the FSR method has been derived and various applications 
to nonperturbative problems have been presented. In the next section, as a 
check of the FSR method, we will obtain perturbative results using the FSR.  

\section{Perturbative expansion from the FSR approach}

\begin{figure}
\begin{center}
\mbox{
   \epsfxsize=1.5in
\epsffile{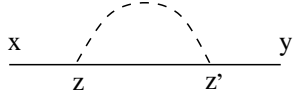} }
\end{center}
\caption{The simplest bubble diagram}
\label{self.fig}
\end{figure}

In this section we will show that perturbation theory results can be obtained 
from the FSR expressions. Let us first consider the perturbative expansion of 
the 1-body self energy in $\chi^2\phi$ interaction. The exact Greens function 
can be expanded in a power series in $g^2$:
\bea
G&=& \biggl\lmoustache_0^\infty\,ds\,\biggl\lmoustache {\cal D}z\,{\rm exp}
\bigl[ -m^2s-\frac{1}{4}\int_0^sd\tau z^2 + \frac{g^2}{2}\int d\tau\int d\tau' \Delta(z(\tau)-z(\tau')\bigr]\nonumber\\
&=&G^{(0)}+G^{(1)}+\cdots.
\eea
The leading contribution ($O(g^0)$) is given by:
\bea
G^{(0)}&=&\biggl\lmoustache_0^\infty\,ds\,\biggl\lmoustache {\cal D}z\,{\rm exp}
\bigl[ -m^2s-\frac{1}{4}\int_0^sd\tau z^2 \bigr],\nonumber\\
&=&\biggl\lmoustache_0^\infty ds \biggl(\frac{N}{4\pi s}\biggr)^{2N} \Pi_{n=1}^{N-1} d^4z_n\,{\rm exp}
\bigl[ -m^2s-\frac{N}{4s}\sum_{i=1}^{N}(z_i-z_{i-1})^2 \bigr].
\eea
Noting that:
\beas
&&\biggl\lmoustache dz_1 dz_2\cdots dz_n\,{\rm exp} \biggl[ -\lambda\{(z_1-x)^2+(z_2-z_1)^2+\cdots+(z_n-y)^2\}\biggr]\\
&&=\biggl[\frac{\pi^n}{(n+1)\lambda^n}\biggr]^{1/2},
\eeas
all $z$ integrals can be performed and the free ($O(g^0)$) propagator is 
found as
\bea
G^{(0)}&=&\biggl\lmoustache_0^\infty \frac{ds}{(4\pi s)^2} \,{\rm exp}\bigl[-m^2 s-\frac{1}{4s}(x-y)^2\bigr]\\
&=&\frac{m}{4\pi^2|x-y|}K_1(m|x-y|),
\eea
which is the free propagator for a massive scalar particle in 3+1d. Now 
let us consider the next to leading order $O(g^2)$ contribution to the 1-body 
propagator. Order $O(g^2)$ term is given by
\beas
G^{(1)}&=&\frac{g^2}{2}\biggl\lmoustache_0^\infty ds\,\biggl\lmoustache {\cal D}z\, \biggl\lmoustache_0^s d\tau\biggl\lmoustache_0^s d\tau'
\,{\rm exp}
\bigl[ -m^2s-\frac{1}{4}\int_0^s d\tau z^2\bigr]\Delta(z(\tau)-z(\tau')).
\eeas
This expression has the following structure
\bea
G^{(1)}&\equiv&\frac{g^2}{2}\biggl\lmoustache_0^\infty ds\biggl\lmoustache_0^s d\tau\biggl\lmoustache_0^s d\tau' f(s,\tau,\tau').
\eea
Using the identity
\bea
\biggl\lmoustache_0^\infty ds\biggl\lmoustache_0^s d\tau \,g(s,\tau)=\biggl\lmoustache_0^\infty ds\biggl\lmoustache_0^\infty d\tau\, g(s+\tau,\tau),
\label{identity}
\eea
one may write
\beas
\biggl\lmoustache_0^\infty ds\biggl\lmoustache_0^s d\tau\biggl\lmoustache_0^\tau d\tau'\,
f(s,\tau,\tau')&=&\biggl\lmoustache_0^\infty ds\biggl\lmoustache_0^\infty d\tau\biggl\lmoustache_0^\infty d\tau'\,f(s+\tau+\tau',\tau+\tau',\tau').
\eeas
Therefore limits of all integrals in $G^{(1)}$ can be extended to infinity
\bea
G^{(1)}&=&g^2\biggl\lmoustache_0^\infty ds\biggl\lmoustache_0^\infty d\tau\biggl\lmoustache_0^\infty d\tau'\,[f(s+\tau+\tau',\tau+\tau',\tau')
\eea
Path integral in $G^{(1)}$ can be split into two regions using:
\bea
({\cal D}z)_{xy}=({\cal D}z)_{xz}\,d^4z\,({\cal D}z)_{zy}.
\eea
Therefore the full expression for $G^{(1)}$ takes the following form
\bea
G^{(1)}&=&g^2\biggl\lmoustache d^4z \biggl\lmoustache d^4z'\,\Delta(z-z')\biggl\lmoustache_0^\infty ds\biggl\lmoustache_0^\infty d\tau\biggl\lmoustache_0^\infty d\tau'\biggl\lmoustache ({\cal D}z)_{xz}({\cal D}z)_{zz'}({\cal D}z)_{z'y}\label{g11}\\
&\times&{\rm exp}\biggl[-m^2(s+\tau+\tau')-\frac{1}{4}\int_0^{\tau'} \dot{z}^2 d\tau''-\frac{1}{4}\int_{\tau'}^{\tau+\tau'} \dot{z}^2 d\tau''-\frac{1}{4}\int_{\tau+\tau'}^{s+\tau+\tau'} \dot{z}^2 d\tau'' \biggr],\nonumber
\eea
where the intermediate boundary conditions are $z\equiv z(\tau')$, and 
$z'\equiv z(\tau+\tau')$. All three path integrals in Eq.~\ref{g11} can now be
integrated to give three free propagators between points $(x-z-z'-y)$  
\bea
G^{(1)}&=&g^2\biggl\lmoustache d^4z \biggl\lmoustache d^4z'\,\Delta(z-z')S(x,z)S(z,z')S(z',y).
\eea
This is the correct perturbative 1-loop bubble expression, shown in 
Fig.~\ref{self.fig}, as expected.
Next let us consider the tadpole diagram.
\subsection{Leading order tadpole diagram from the FSR approach }  
\begin{figure}
\begin{center}
\mbox{
   \epsfxsize=1.5in
\epsffile{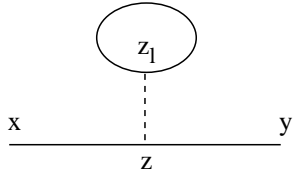} }
\end{center}
\caption{The leading order connected 1-loop diagram}
\label{tadpole.fig}
\end{figure}

The tadpole diagram exists only after unquenching. Consider the 1-body 
propagator,
\beas
G(y|x)&=&N\bigg\lmoustache {\cal D}\phi\, ({\rm det}S)\,S(x,y)\,e^{-S[\phi]}.
\eeas
In the quenched approximation det(S) is set equal to one. Here let us consider 
the next to leading order contribution to the quenched approximation. We can
make the following expansion
\beas
{\rm det}S&=&{\rm exp}\bigl[{\rm tr}({\rm log}S)\bigr]\\
&\simeq&1+{\rm tr}({\rm log}S)+ \frac{  \left[{\rm tr}({\rm log}S)\right]^2}{2}+\cdots,
\eeas
where the trace is defined by
\beas
{\rm tr}({\rm log}S)&=&\biggl\lmoustache d^4x \left<x| {\rm log}S |x\right>.
\eeas
Note that the trace is infinite and requires regularization, but this 
does not effect our discussion regarding the derivation of the perturbative 
result. Ordinary propagator has a path integral representation,
\beas
S&=&\bigg\lmoustache_0^\infty ds\,\bigg\lmoustache\,{\cal D}z\,\,{\rm exp}\biggl[-sm^2-\frac{1}{4}\int_0^s d\tau\, \dot{z}^2(\tau)-g\int_0^s d\tau\, \phi(z(\tau))\,\,\biggr].
\eeas
The logarithm of the propagator can also be expressed in the form of a path 
integral
\beas
{\rm log}S=\bigg\lmoustache_0^\infty \frac{ds}{s}\,\bigg\lmoustache\,{\cal D}z\,\,{\rm exp}\biggl[-sm^2-\frac{1}{4}\int_0^s d\tau\, \dot{z}^2(\tau)-g\int_0^s d\tau\, \phi(z(\tau))\,\,\biggr].
\eeas
Therefore the diagram with 1-loop connected to the propagator can be written as
\beas
G(x,y)&=&\biggl\lmoustache_0^\infty ds\biggl\lmoustache_0^\infty \frac{ds_l}{s_l}\biggl\lmoustache ({\cal D}z)_{xy}\biggl\lmoustache ({\cal D}z_l)\, {\rm exp}\biggl[-m^2s-m^2s_l\\
&-&\int_0^s\dot{z}^2\,d\tau-\int_0^{s_l}\dot{z}^2\,d\tau+g^2\int_0^s\oint_0^{s_l}\,d\tau\, d\tau_l\,\Delta(z(\tau)-z_l(\tau_l))\biggr].
\eeas
The next to leading order contribution to the 1-loop-1-particle connected
propagator is given by
\bea
G(x,y)&=&g^2\biggl\lmoustache_0^\infty ds\biggl\lmoustache_0^\infty \frac{ds_l}{s_l}\biggl\lmoustache ({\cal D}z)_{xy}\biggl\lmoustache ({\cal D}z_l)\,
\biggl\lmoustache_0^s\,d\tau \biggl\lmoustache_0^{s_l}\, 
d\tau_l\,\Delta(z(\tau)-z_l(\tau_l))\nonumber\\
&&\times\,{\rm exp}\biggl[-m^2s-m^2s_l-\int_0^s\dot{z}^2\,d\tau-\int_0^{s_l}
\dot{z}^2\,d\tau\biggr],
\eea
where the loop trajectory $z_l$ is a circular trajectory and therefore 
has no fixed initial or final coordinates. 
Using the identity Eq.~\ref{identity} and splitting the path integral 
as before one obtains
\beas
&&G(x,y)=g^2 \biggl\lmoustache_0^\infty ds\biggl\lmoustache_0^\infty \frac{ds_l}{s_l+\tau_l}\biggl\lmoustache_0^\infty\,d\tau \biggl\lmoustache_0^\infty\,d\tau_l\,{\rm exp}\biggl[ -m^2(s+s_l+\tau+\tau_l)\biggr]\\
&&\times\biggl\lmoustache({\cal D}z)_{xz(\tau)} \,d^4z(\tau)\,({\cal D}z)_{z(\tau)y}\biggl\lmoustache\,d^4z_0\,({\cal D}z_l)_{z_0z_l(\tau_l)} \,d^4z_l(\tau_l)\,({\cal D}z_l)_{z_l(\tau_l)z_0}\\
&&\times \Delta(z(\tau)-z_l(\tau_l))\,{\rm exp}\biggl[-\int_0^\tau \dot{z}^2\, d\tau -\int_\tau^{s+\tau}\dot{z}^2\, d\tau
-\int_0^{\tau_l}\dot{z}_{l}^2\, d\tau - \int_{\tau_l}^{s_l+\tau_l} 
\dot{z}^2_{l}\,d\tau \biggr],
\eeas
where the boundary conditions are 
\bea
&&z(0)=x,\ \  z(\tau)=z,\ \  z(s+\tau)=y,\\
&&z_{l}(0)=z_{l}(s_l+\tau_l)=z_0,\ \  z_{l}(\tau_l)=z_{l}.
\eea
Let us note that 
\bea
\frac{{\rm exp}[-m^2(s+\tau)]}{s+\tau}&=&\biggl\lmoustache_{m^2}^\infty dm^2 \,{\rm exp}[-m^2(s+\tau)].
\eea
With this replacement all path integrals can be evaluated easily, as done 
earlier in the 1-body self energy calculation, and the 1-loop Greens function 
to $O(g^2)$ reduces to 
\beas
G(x,y)=\\
&&g^2 \biggl\lmoustache d^4z \biggl\lmoustache d^4z_l\, S(x,z) S(z,y) \Delta(z-z_l) \biggl\lmoustache_{m^2}^\infty dm^2\, \biggl\lmoustache d^4z_0\, S_m(z_0,z_l)S_m(z_l,z_0).
\eeas
$z_0$ and $m^2$ integrals can be performed by noting that:
\bea
\biggl\lmoustache d^4z_0\, S_m(z_0,z)S_m(z,z_0)&=&-\frac{\partial}{\partial m^2} S_m(z,z),\\
\lim_{m^2\rightarrow \infty}S_m(x,y)&=&0.
\eea
Therefore the 1-loop diagram to $O(g^2)$ is found as 
\bea
G(x,y)&=&g^2 \biggl\lmoustache d^4z \biggl\lmoustache d^4z_l\, S(x,z) S(z,y) \Delta(z-z_l) S(z_l,z_l),
\eea
which is the expected result from the perturbation theory, as shown in 
Fig.~\ref{tadpole.fig}.
This concludes the discussion of perturbative results within the FSR method.
It should be clear that one may extend this discussion to higher order 
diagrams and obtain the correct perturbation theory results. It should 
be noted that unquenching to all orders in the FSR approach is
numerically not feasable. This is related to the fact that the FSR approach 
relies on the discretization of trajectories. Every loop involved in 
the calculation represents a new  discretized trajectory. Therefore 
inclusion of all loops is practically not possible. However one maybe be able 
to extract information about the effect of unquenching by making an expansion 
in the number of loops, that is by introducing loops order by order. More 
work needs to be done on this topic.

\section{Conclusions}
In these lectures the FSR representation has been introduced with various 
applications to scalar field theories. It has been shown that the FSR is
an efficient and rigorous method for doing nonperturbative calculations in 
field theory. The FSR approach uses a covariant path integral representation 
for the trajectories of particles. Reduction of field theoretical path 
integrals to path integrals involving particle trajectories reduces the 
dimensionality of the problem and the associated computational cost. The FSR 
uses a space-time continuum. There are no boundaries in space-time and 
rotational symmetry is respected. 

Applications of the FSR approach to 1 and 2-body problems in particular shows 
that uncontrolled approximations in field theory may lead to significant 
deviations from the correct result. Results presented here indicate that 
the ladder approximation for the 2-body bound state problem, and the rainbow 
approximation for the 1-body problem are both poor approximations. In both 
cases the crossed diagrams (such as crossed ladders) play an essential role.

\vspace{1cm}
\noindent {\bf Acknowledgements:}
I would like to thank the organizers of the 13th Indian-Summer School in 
Prague for inviting me to give a series of lectures, and for giving me 
a chance to experience the culture and the history of this beautiful city.
This work was supported in part by the US Department of Energy under grant 
No.~DE-FG02-97ER41032. Author thanks F. Gross and J. Tjon for useful 
discussions.

\end{document}